\documentclass[aps,prl, twocolumn,groupedaddress]{revtex4-1}
\usepackage{bm}
\usepackage{graphicx}
\usepackage{epstopdf}
\usepackage{amsmath}

\begin{document}


\title{CeB$_6$ macroscopically revisited}


\author{M. Amara}
\email{mehdi.amara@grenoble.cnrs.fr}
\altaffiliation{Universit\'e Joseph-Fourier, Grenoble}

\author{R.-M. Gal\'era}
\affiliation{Institut N\'eel, CNRS et Universit\'e Joseph Fourier, BP 166, F-38042 Grenoble Cedex 9, France}


\date{\today}

\begin{abstract}
Magnetic susceptibility anisotropy and high sensitivity magnetostriction measurements are used to investigate the symmetry of CeB$_6$ ordered states. The antiferromagnetic state is confirmed as tetragonal, but no deviation from the cubic symmetry is observed in the so-called antiferroquadrupolar phase, where only volume effects are detected. In this phase, the temperature dependence of the strain field-susceptibilities is typical of non-ordered quadrupoles. Moreover, while an antiferroquadruplar order should be cubic, this symmetry is incompatible with the $\langle\frac{1}{2}\frac{1}{2}\frac{1}{2} \rangle$ ordering wave-vector. The antiferroquadrupolar description of CeB$_6$ phase II is clearly inconsistent and an alternative model, based on a unidimensional representation of the cube, has to be sought for. 

\end{abstract}


\maketitle

\section{}
\begin{figure}
\includegraphics[width=\columnwidth]{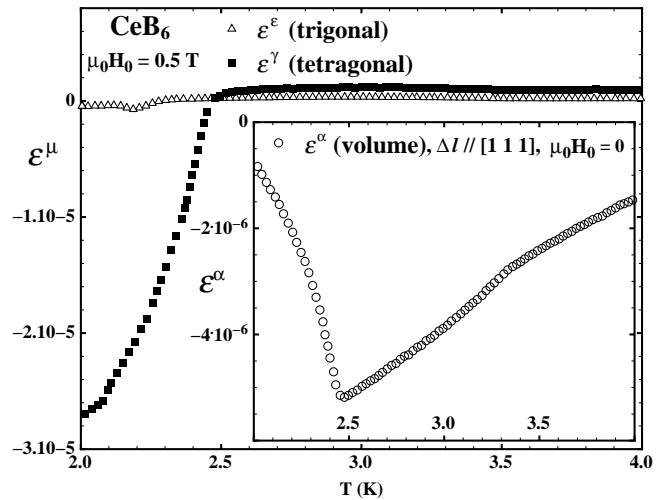}
\caption{\label{Epsilon(T)} Temperature variation of the $\gamma$ and $\varepsilon$ strains, as obtained from measurements under a 0.5 T magnetic field. The inset shows the volume $\alpha$ strain deduced from the length of [111], in zero field.}
\end{figure}

\begin{figure}
\includegraphics[width=\columnwidth]{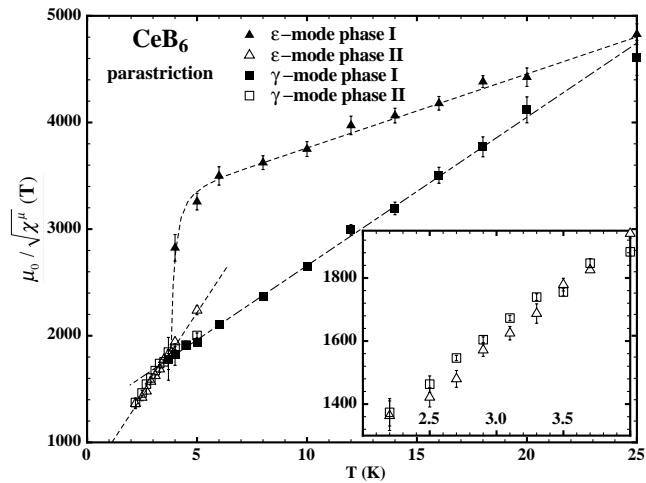}
\caption{\label{CondParastrict} Thermal dependencies of $1/\sqrt{\chi^\gamma}$ and $1/\sqrt{\chi^\varepsilon}$ in the range of phase I an II. The lines are just guides to the eye and the inset details the data for phase II.}
\end{figure}

In rare-earth compounds, the degrees of freedom of the 4$f$ ions are collectively involved at low temperature to form ordered states. The most common kind of order is the magnetic one that exists in a number of varieties. The orbital order is a less common situation, where only the orbital degeneracy of the 4$f$ ion is involved, without the stabilization of magnetic moments. It requires the orbital degeneracy to survive the crystal-field effect, which means the 4$f$ ions sites are of high symmetry. As the highest symmetry is that of the octaedra, cubic systems are the  most favorable scene for the development of orbital related effects \cite{MorinSchmitt1990}. There, the lifting of the degeneracy is equivalent to the emergence of 4$f$ electric quadrupole moments and the associated order are referred to as "quadrupolar". By analogy with the ferromagnetism, the order that doesn't change the crystal periodicity is called ferroquadrupolar. One can also consider a state where the 4$f$ quadrupoles are changing from site to site, thus defining an "antiferroquadrupolar" order (AFQ), analog to the antiferromagnetism. For more than two decades, the most cited example of such an antiferroquadrupolar order has been CeB$_6$ phase II.\\
This system crystallizes within the CaB$_6$-type structure and displays complex magnetic phase diagrams \cite{Kawakami1980, Fujita1980}. CeB$_6$ orders antiferromagnetically at T$_N$ = 2.4 K (phase III) and its most intriguing feature is the occurrence of a non-magnetic state (phase II) below T$_Q$= 3.2 K, before entering the paramagnetic state (phase I). In phase II, a magnetic field induced $\langle\frac{1}{2}\frac{1}{2}\frac{1}{2} \rangle$ antiferromagnetism is detected via neutron diffraction \cite{Effantin1985}. Latter X-ray scattering experiments \cite{Nakao2001, Yakhou2001, Matsumura2009, Tanaka2004} have confirmed a non-magnetic $\langle\frac{1}{2}\frac{1}{2}\frac{1}{2}\rangle$ wave-vector, even in zero field.\\
Inelastic neutron scattering \cite{Zirngiebl1984} results are consistent with a crystal-field splitting of the Ce$^{3+}$, J = 5/2, multiplet resulting in a $\Gamma_8^+$ quadruplet ground state well separated (530 K) from the $\Gamma_7^+$ doublet. As this ground state has an orbital degeneracy, Effantin \textit{et al.} \cite{Effantin1985} interpreted the puzzling properties of phase II as resulting from an AFQ state.\\
Up to now, no solidly based model has been given for the microscopic organization of this phase. Some authors suggest a $\Gamma_3^+$ AFQ ordering \cite{Luthi1984} while, for others, it is of the $\Gamma_5^+$ type. The recurrent scheme is a $\Gamma_5^+$ AFQ state \cite{Sakai1997}, where a single quadrupolar component, $P_{xy}$, is propagated by the $[\frac{1}{2}\frac{1}{2}\frac{1}{2} ]$ wave-vector. This model doesn't preserve the cubic symmetry and, due to the coupling of the quadrupoles with the lattice, a $\Gamma_3^+$ ferroquadrupolar component and a macroscopic distortion should simultaneously develop. Facing such inconsistencies and a confusing mass of experimental data, we thought that a crucial clarification would come from an accurate investigation of phase II symmetry.\\
This can be achieved from macroscopic experiments on a single crystal. For a rare-earth based system, magnetic measurements are first to consider. In cubic crystals, an accurate determination of the first-order magnetic susceptibility can reveal a symmetry lowering : any deviation from the cubic symmetry will result in an anisotropic susceptibility. Due to the magnetoelastic coupling, the lattice necessarily reflects a symmetry lowering. The detection of the associated strain can be very sensitive, capacitance dilatometer being able to detect relative changes as small as 10$^{-8}$. The difficulty, when investigating a sample undergoing a symmetry lowering, is that it divides into domains, mixing their properties. This can be solved using a magnetic field of appropriate direction, in order to coerce the sample to a single domain state. If the anisotropy of the susceptibility has been previously determined, the optimal field direction is easily defined. All the reported experiments where made using single crystals in the shape of rectangular platelets (6x4x1 mm$^3$) previously used for neutron scattering experiments \cite{Plakhty2005}.\\
\begin{figure}
\includegraphics[width=\columnwidth]{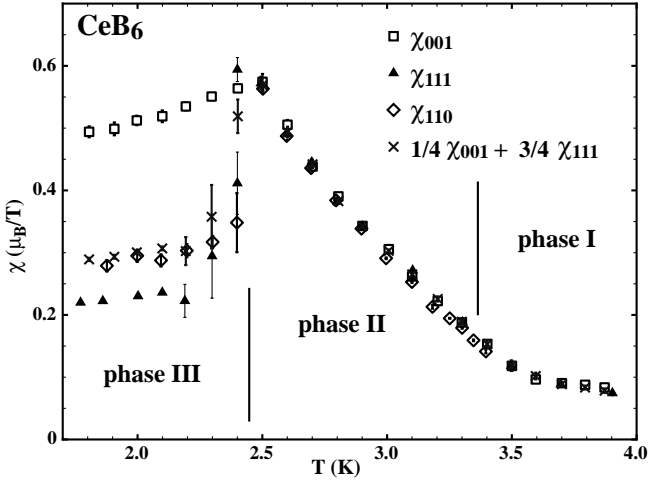}
\caption{\label{chiplot} Magnetic susceptibilities of CeB$_6$ along the three high symmetry directions of the cube, as deduced from Arrott's plots in descending field.}
\end{figure}
The investigation of the magnetic susceptibility is based on magnetization measurements along the three main cubic directions ([001], [110] and [111]) in fields up to 8 T. The susceptibilities, $\chi_{001}$, $\chi_{110}$ and $\chi_{111}$, are extracted using the Arrott method \cite{Arrott1957} applied to the descending field data : only the field selected domains should contribute to the magnetization. Then, considering the tetragonal and trigonal symmetry lowerings \cite{Amara2010} cases, the measured susceptibility can be related to the normal susceptibilities \cite{Amara2010}, $\chi_{\|}$ and $\chi_{\bot}$, respectively parallel and perpendicular to the preserved, fourfold or threefold, axis of the system. One can thus identify the symmetry both via hierarchical and quantitative relations between $\chi_{001}$, $\chi_{110}$ and $\chi_{111}$.\\
Fig. \ref{chiplot} shows the thermal dependence of these three susceptibilities at temperatures below 4.5 K.
The transition between phases I and II is difficult to discern, both phases displaying a paramagnetic decrease of the susceptibility with the temperature. Within the precision of the measurements, typically 2\%, no difference between the three susceptibilities and, therefore, no deviation from the cubic symmetry can be detected.
In the antiferromagnetic state, the three susceptibilities are well separated, with a dominant $\chi_{001}$, an intermediate $\chi_{110}$ and a minimal $\chi_{111}$. This sequence is expected in case of a tetragonal symmetry lowering, with a maximum susceptibility along the fourfold axis ($\chi_{\|}>\chi_{\bot}$). In quantitative terms, this translates into $\chi_{110}=\frac{1}{4}\chi_{001}+\frac{3}{4}\chi_{111}$, which is precisely what is observed on Fig. \ref{chiplot}. This is consistent with the multiaxial antiferromagnetic models \cite{Effantin1985} with magnetic moments perpendicular to the fourfold axis.\\
\begin{figure}
\includegraphics[width=\columnwidth]{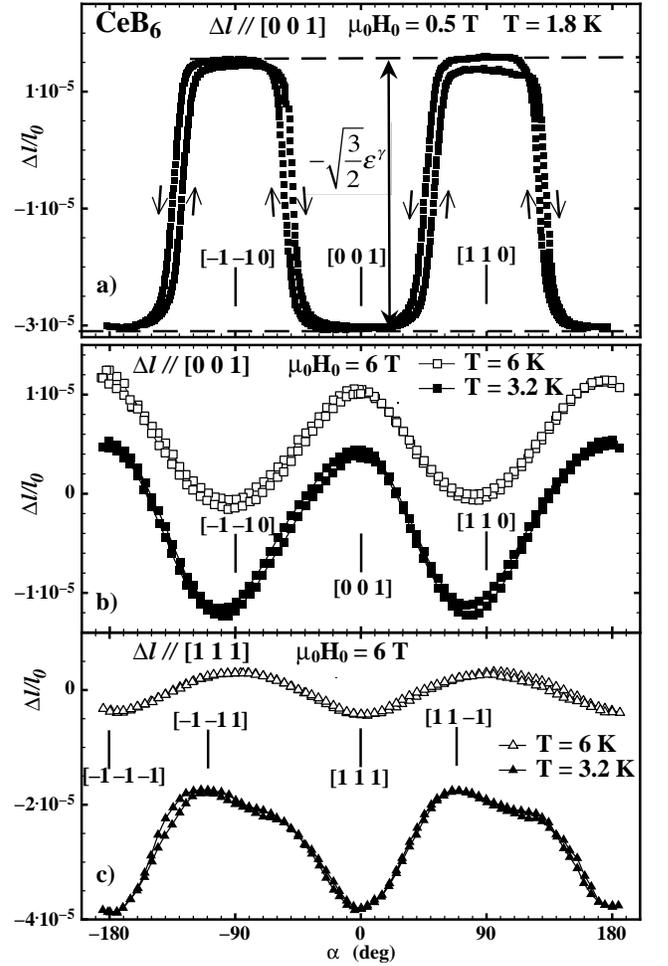}
\caption{\label{VarAngl} Relative changes in length along [001] (a and b) and [111] (c) while rotating a constant field in the (110) plane in the ranges of phase III (a), phase II (b and c) and phase I (b and c). The zero angle $\alpha$ correspond to a field aligned with the measured crystal direction. The vertical lines indicate high symmetry directions in the (110) plane.}
\end{figure}
Beyond the susceptibility analysis, one can further track a symmetry lowering using a high sensitivity dilatometer. For this purpose, we used the magnetostriction setup of the "Institut N\'eel", which is based on a capacitance cell that can be rotated in a 0-6 T horizontal magnetic field, in the 2-300 K temperature range. The sensitivity, in terms of relative change in length, is better than 10$^{-7}$. The alignment of the sample inside the setup is defined by the the choice of the crystal horizontal plane, which contains $[mnp]$, the probed crystal's direction, and $[ijk]$ the direction of the applied field. For a reference length $l_0$, a current length $l$, the relative change in length, and its geometrical conditions, can be noted as :
$ {}^{mnp}\lambda _{ijk}  = \Delta l/l_0 = (l-l_0)/l_0$.\\
In the following, the default reference length $l_0$ is taken at T = 5 K and in zero field. All the striction phenoma are analyzed in terms of cubic normal strain modes \cite{Callen1963}. For the most common cases of a tetragonal or trigonal symmetry lowering, the strain ${}^{mnp}\lambda$ of a single domain depends on three quantities: $\varepsilon^\alpha$, the volume strain, $\varepsilon^\gamma$, the tetragonal strain, and the trigonal one $\varepsilon^\varepsilon$.\\
To detect a symmetry lowering, one can take advantage of the adjustable field direction : rotating the sample in a constant field will successively select different domains, inducing changes in the probed sample length. Fig. \ref{VarAngl} shows such measurements while rotating the field in the (110) plane, where all high symmetry axes of the cube are represented. In Fig. \ref{VarAngl} (a) and \ref{VarAngl} (b), the probed sample direction is [001], which is ideal for detecting a tetragonal mode. Fig. \ref{VarAngl} (a) shows what happens in the antiferromagnetic phase, under a 0.5 T applied field : spectacular, square like, changes in the sample length are observed. This is expected in case of a tetragonal symmetry lowering with maximum susceptibility along the fourfold axis : one goes from a single domain with [001] fourfold axis, for low ${\alpha}$ values, to a $\alpha = \pm 90 ^\circ $ state where the two other domains are selected, with fourfold axes [010] and [100]. The amplitude of the jump is directly related to the $\varepsilon^\gamma$ spontaneous strain of the antiferromagnetic state. Fig. \ref{VarAngl} (b) gives the angular dependence under the maximum field (6 T) in phases II (T = 3.2 K) and phase I (T = 6 K). These curves show no anomaly but a continuous, sine-type, change in the [001] length, which is typical of the paramagnetic state. The twisted phase II curve is an effect of higher-order susceptibilities in conjunction with a slight sample's misorientation. Fig. \ref{VarAngl} (c) shows the angular variations, in phases II and I, of the [111] length, which is adapted to detect a trigonal mode. Here again, the continuous, undulating curves give no evidence of domains. Higher-order effects are clearly present in phase II, with a resulting double bump around $\alpha = 90 ^\circ $.\\
The temperature dependence of the strain modes has been derived from a sequence of ${}^{mnp}\lambda _{ijk}$ measurements under a constant (0.5 T) field (Fig. \ref{Epsilon(T)}). It is only within the antiferromagnetic range that a spontaneous, tetragonal, strain is detected. The corresponding small accident on the trigonal curve is the result of an imperfect sample alignment : a small fraction of the tetragonal strain is detected. In absence of a trigonal mode, the volume effect $\varepsilon^\alpha$ (inset of Fig. \ref{Epsilon(T)}) can be safely derived from the length of a threefold axis : both transitions are visible, with a more pronounced thermal expansion in phase II, while phase III shows a substantial negative effect. This agrees with previous measurements by Sera \textit{et al.} \cite{Sera1988}.\\
The magnetic and magnetostriction data are consistent with a cubic symmetry in phases II and I. In this cubic context, the magnetostriction setup can be used to probe the magnetoelastic susceptibilities of the system with respect to a field enforced tetragonal (field along a fourfold axis) or trigonal (field along a threefold axis) symmetry lowering. This is the parastriction analysis of the paramagnetic state magnetoelastic properties \cite{Morin1980A}. At the lowest order, the strains vary quadratically with the applied field : $\varepsilon^{\mu} = \chi^{\mu} H^2$, where $\chi^{\mu}$ is a magnetoelastic susceptibility and $\mu$ states for $\gamma$ or $\varepsilon$. In the harmonic treatment of 4$f$ magnetoelasticity, $\chi^\mu$ is proportional to the quadrupolar field-susceptibility $\chi^{\mu}_Q$ of the same symmetry. $\chi^\mu(T)$ can be deduced by linearizing the constant temperature strain as function of the squared applied field.
\begin{figure}
\includegraphics[width=\columnwidth]{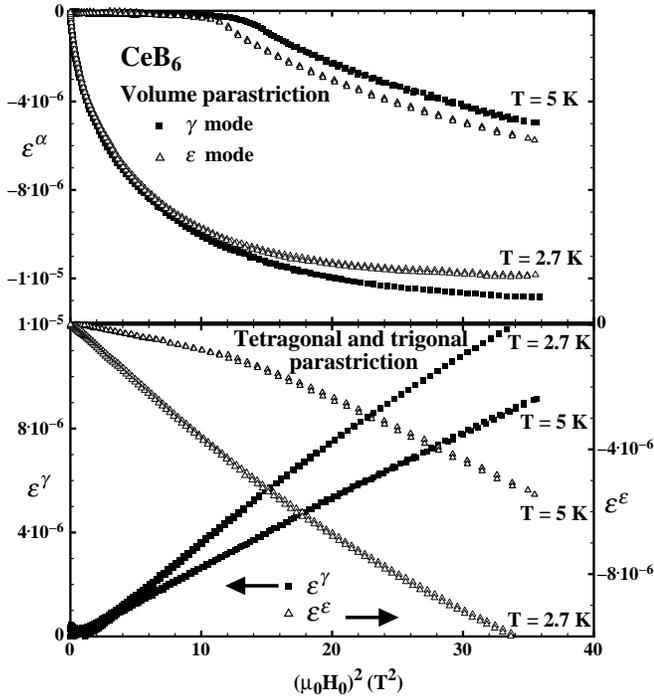}
\caption{\label{Parastrict} Examples of parastriction curves in phases I and II, at T = 2.7 and 5 K, measured for the $\gamma$ and $\varepsilon$ symmetry lowering modes. Upper part :  squared field dependence of the volume mode.  Lower part : the tetragonal and trigonal strains.}
\end{figure}
Fig. \ref{Parastrict} shows examples of such curves in phase II (T = 2.7 K) and at the field crossing from phase I into phase II (T = 5 K). It appears that the latter has a surprisingly large volume dependence on the applied field (Fig. \ref{Parastrict}, upper part). On the $\varepsilon^\gamma$ parastriction curve, the transition from phase I to II is almost unnoticeable, whereas for the $\varepsilon$ mode, the negative slope is amplified in phase II at T = 5 K (Fig. \ref{Parastrict}, lower part). No hysteresis is observed in these curves, which confirms the absence of domain effects. In both phases, the absolute value of the susceptibility decreases with the temperature. This is expected if the degrees of freedom responsible for the field induced strains are not ordered. Quadrupoles are considered as the dominant actors in the parastriction of cubic rare-earth compounds, which results in a quadratic Curie-like behaviour for $\chi^\mu(T)$. To test this behaviour, $1/\sqrt{\chi^\mu}$ is represented as a function of T in Fig. \ref{CondParastrict}. In both phases, the curves seem indeed to vary linearly with the temperature. In phase II, the data for the $\gamma$ and $\varepsilon$ modes are almost superimposed. As these measurements mix elastic and magnetic properties, such an identity for different symmetry modes is fortuitous.\\
No symmetry lowering is detected in phase II, which discards non cubic models as the one of Ref. \cite{Sakai1997}. While a true AFQ state needs to be cubic, the parastriction results give evidence against the ordering of the quadrupoles. This is consistent with the elastic constants measurements \cite{Luthi1984}, probes of the quadrupoles stress-susceptibilities, which, at least for the trigonal mode, behave as if quadrupoles where not ordered. Nevertheless, we can deepen the analysis by specifying the characteristics of a cubic AFQ order. As the transition at T$_Q$ is second order, the ordered state should involve a single quadrupolar representation, $\Gamma_3^+$ ($\gamma$) or $\Gamma_5^+$ ($\varepsilon$), for a single wave-vector star. To obtain a high symmetry structure, stable down to 0 K (above 2.5 Tesla), all the rare-earth ions should be equivalent by symmetry.\\
Considering the $\Gamma_3^+$ representation, the emergence of a 4$f$ quadrupole is equivalent to a local tetragonal symmetry lowering : in a polar representation, the 4$f$ electronic distribution is stretched (or squeezed) along one of the fourfold axes of the cube. There are then three symmetry equivalent $\Gamma_3^+$ modes for the 4$f$ ion.
As for the $\Gamma_5^+$ representation, the quadrupoles develop with a local trigonal symmetry lowering, which means that there are four possible 4$f$ modes, one for each threefold axis. The cubic symmetry is preserved if all the modes, three for $\Gamma_3^+$ or four for $\Gamma_5^+$, are equally represented in the cubic AFQ cell. The smallest cell one can consider is built from the 8 sites at the vertexes of a cube, which is enough to describe structures with $\langle\frac{1}{2}00\rangle$, $\langle\frac{1}{2}\frac{1}{2}0\rangle$ and $\langle\frac{1}{2}\frac{1}{2}\frac{1}{2}\rangle$ wave-vectors. On this cell, it is clearly impossible to equally represent the three $\Gamma_3^+$ modes : there exists no cubic $\Gamma_3^+$ AFQ structure for these wave-vectors. This is not the case for $\Gamma_5^+$ and cubic AFQ states are possible for the $\langle\frac{1}{2}00\rangle$ and $\langle\frac{1}{2}\frac{1}{2}0\rangle$ wave-vectors. However, this is strictly impossible for $\langle\frac{1}{2}\frac{1}{2}\frac{1}{2}\rangle$, the only wave-vector that is detected in phase II. This star reduces to a single branch which defines only two 4$f$ modes, with opposite sign. An object that alternates in sign from site to site, while preserving the cubic symmetry, necessarily belongs to a unidimensional representation of the cube. Among the even $O_h$ group representations, only $\Gamma_2^+$ allows a change in sign. To find an electric multipole transforming according to $\Gamma_2^+$, one has to go up to 64 poles, which means a sixth-order operator in $\bm{J}$ components. For J = 5/2, these hexacontatetrapoles cancel : no 4$f$ electric multipole can describe the cubic $\langle\frac{1}{2}\frac{1}{2}\frac{1}{2}\rangle$ ordering of CeB$_6$.\\
To solve this puzzle, one has to consider alternative objects, transforming according to one of the unidimensional representations of $O_h$. In a cage system, as suggested by Kasuya \cite{Kasu1998}, one may consider the dynamic decentering of the 4$f$ ion, already present above T$_Q$ as a Jahn-Teller effect. Below T$_Q$ there might be a symmetry breaking in the spatial distribution of the Ce ion, that would alternate from site to site. In contrast with the 4$f$ electronic configuration, the related electric multipoles can be of odd parity : in addition to $\Gamma_2^+$, they may also transform according to the $\Gamma_1^-$ or $\Gamma_2^-$ representations.\\
The authors are indebted to Dr L.P. Regnault, who provided us with the sample and critically read the manuscript.


\begin{thebibliography}{18}%
\makeatletter
\providecommand \@ifxundefined [1]{%
 \@ifx{#1\undefined}
}%
\providecommand \@ifnum [1]{%
 \ifnum #1\expandafter \@firstoftwo
 \else \expandafter \@secondoftwo
 \fi
}%
\providecommand \@ifx [1]{%
 \ifx #1\expandafter \@firstoftwo
 \else \expandafter \@secondoftwo
 \fi
}%
\providecommand \natexlab [1]{#1}%
\providecommand \enquote  [1]{``#1''}%
\providecommand \bibnamefont  [1]{#1}%
\providecommand \bibfnamefont [1]{#1}%
\providecommand \citenamefont [1]{#1}%
\providecommand \href@noop [0]{\@secondoftwo}%
\providecommand \href [0]{\begingroup \@sanitize@url \@href}%
\providecommand \@href[1]{\@@startlink{#1}\@@href}%
\providecommand \@@href[1]{\endgroup#1\@@endlink}%
\providecommand \@sanitize@url [0]{\catcode `\\12\catcode `\$12\catcode
  `\&12\catcode `\#12\catcode `\^12\catcode `\_12\catcode `\%12\relax}%
\providecommand \@@startlink[1]{}%
\providecommand \@@endlink[0]{}%
\providecommand \url  [0]{\begingroup\@sanitize@url \@url }%
\providecommand \@url [1]{\endgroup\@href {#1}{\urlprefix }}%
\providecommand \urlprefix  [0]{URL }%
\providecommand \Eprint [0]{\href }%
\providecommand \doibase [0]{http://dx.doi.org/}%
\providecommand \selectlanguage [0]{\@gobble}%
\providecommand \bibinfo  [0]{\@secondoftwo}%
\providecommand \bibfield  [0]{\@secondoftwo}%
\providecommand \translation [1]{[#1]}%
\providecommand \BibitemOpen [0]{}%
\providecommand \bibitemStop [0]{}%
\providecommand \bibitemNoStop [0]{.\EOS\space}%
\providecommand \EOS [0]{\spacefactor3000\relax}%
\providecommand \BibitemShut  [1]{\csname bibitem#1\endcsname}%
\let\auto@bib@innerbib\@empty
\bibitem [{\citenamefont {Morin}\ and\ \citenamefont
  {Schmitt}(1990)}]{MorinSchmitt1990}%
  \BibitemOpen
  \bibfield  {author} {\bibinfo {author} {\bibfnamefont {P.}~\bibnamefont
  {Morin}}\ and\ \bibinfo {author} {\bibfnamefont {D.}~\bibnamefont
  {Schmitt}},\ }\enquote {\bibinfo {title} {Quadrupolar interactions and
  magnetoelastic effects in rare earth intermetallic compounds},}\ \ (\bibinfo
  {publisher} {Elsevier Science},\ \bibinfo {year} {1990})\ Chap.~\bibinfo
  {chapter} {1}, pp.\ \bibinfo {pages} {1--132}\BibitemShut {NoStop}%
\bibitem [{\citenamefont {Kawakami}\ \emph {et~al.}(1980)\citenamefont
  {Kawakami}, \citenamefont {Kunii}, \citenamefont {Komatsubara},\ and\
  \citenamefont {Kasuya}}]{Kawakami1980}%
  \BibitemOpen
  \bibfield  {author} {\bibinfo {author} {\bibfnamefont {M.}~\bibnamefont
  {Kawakami}}, \bibinfo {author} {\bibfnamefont {S.}~\bibnamefont {Kunii}},
  \bibinfo {author} {\bibfnamefont {T.}~\bibnamefont {Komatsubara}}, \ and\
  \bibinfo {author} {\bibfnamefont {T.}~\bibnamefont {Kasuya}},\ }\href
  {\doibase DOI: 10.1016/0038-1098(80)90928-X} {\bibfield  {journal} {\bibinfo
  {journal} {Solid State Communications}\ }\textbf {\bibinfo {volume} {36}},\
  \bibinfo {pages} {435 } (\bibinfo {year} {1980})}\BibitemShut {NoStop}%
\bibitem [{\citenamefont {Fujita}\ \emph {et~al.}(1980)\citenamefont {Fujita},
  \citenamefont {Suzuki}, \citenamefont {Komatsubara}, \citenamefont {Kunii},
  \citenamefont {Kasuya},\ and\ \citenamefont {Ohtsuka}}]{Fujita1980}%
  \BibitemOpen
  \bibfield  {author} {\bibinfo {author} {\bibfnamefont {T.}~\bibnamefont
  {Fujita}}, \bibinfo {author} {\bibfnamefont {M.}~\bibnamefont {Suzuki}},
  \bibinfo {author} {\bibfnamefont {T.}~\bibnamefont {Komatsubara}}, \bibinfo
  {author} {\bibfnamefont {S.}~\bibnamefont {Kunii}}, \bibinfo {author}
  {\bibfnamefont {T.}~\bibnamefont {Kasuya}}, \ and\ \bibinfo {author}
  {\bibfnamefont {T.}~\bibnamefont {Ohtsuka}},\ }\href {\doibase DOI:
  10.1016/0038-1098(80)90900-X} {\bibfield  {journal} {\bibinfo  {journal}
  {Solid State Communications}\ }\textbf {\bibinfo {volume} {35}},\ \bibinfo
  {pages} {569 } (\bibinfo {year} {1980})}\BibitemShut {NoStop}%
\bibitem [{\citenamefont {Effantin}\ \emph {et~al.}(1985)\citenamefont
  {Effantin}, \citenamefont {Rossat-Mignod}, \citenamefont {Burlet},
  \citenamefont {Bartholin}, \citenamefont {Kunii},\ and\ \citenamefont
  {Kasuya}}]{Effantin1985}%
  \BibitemOpen
  \bibfield  {author} {\bibinfo {author} {\bibfnamefont {J.}~\bibnamefont
  {Effantin}}, \bibinfo {author} {\bibfnamefont {J.}~\bibnamefont
  {Rossat-Mignod}}, \bibinfo {author} {\bibfnamefont {P.}~\bibnamefont
  {Burlet}}, \bibinfo {author} {\bibfnamefont {H.}~\bibnamefont {Bartholin}},
  \bibinfo {author} {\bibfnamefont {S.}~\bibnamefont {Kunii}}, \ and\ \bibinfo
  {author} {\bibfnamefont {T.}~\bibnamefont {Kasuya}},\ }\href@noop {}
  {\bibfield  {journal} {\bibinfo  {journal} {J. Magn. Magn. Mater.}\ }\textbf
  {\bibinfo {volume} {47-48}},\ \bibinfo {pages} {145} (\bibinfo {year}
  {1985})}\BibitemShut {NoStop}%
\bibitem [{\citenamefont {Nakao}\ \emph {et~al.}(2001)\citenamefont {Nakao},
  \citenamefont {ichi Magishi}, \citenamefont {Wakabayashi}, \citenamefont
  {Murakami}, \citenamefont {Koyama}, \citenamefont {Hirota}, \citenamefont
  {Endoh},\ and\ \citenamefont {Kunii}}]{Nakao2001}%
  \BibitemOpen
  \bibfield  {author} {\bibinfo {author} {\bibfnamefont {H.}~\bibnamefont
  {Nakao}}, \bibinfo {author} {\bibfnamefont {K.}~\bibnamefont {ichi Magishi}},
  \bibinfo {author} {\bibfnamefont {Y.}~\bibnamefont {Wakabayashi}}, \bibinfo
  {author} {\bibfnamefont {Y.}~\bibnamefont {Murakami}}, \bibinfo {author}
  {\bibfnamefont {K.}~\bibnamefont {Koyama}}, \bibinfo {author} {\bibfnamefont
  {K.}~\bibnamefont {Hirota}}, \bibinfo {author} {\bibfnamefont
  {Y.}~\bibnamefont {Endoh}}, \ and\ \bibinfo {author} {\bibfnamefont
  {S.}~\bibnamefont {Kunii}},\ }\href {\doibase 10.1143/JPSJ.70.1857}
  {\bibfield  {journal} {\bibinfo  {journal} {Journal of the Physical Society
  of Japan}\ }\textbf {\bibinfo {volume} {70}},\ \bibinfo {pages} {1857}
  (\bibinfo {year} {2001})}\BibitemShut {NoStop}%
\bibitem [{\citenamefont {Yakhou}\ \emph {et~al.}(2001)\citenamefont {Yakhou},
  \citenamefont {Plakhty}, \citenamefont {Suzuki}, \citenamefont {Gavrilov},
  \citenamefont {Burlet}, \citenamefont {Paolasini}, \citenamefont {Vettier},\
  and\ \citenamefont {Kunii}}]{Yakhou2001}%
  \BibitemOpen
  \bibfield  {author} {\bibinfo {author} {\bibfnamefont {F.}~\bibnamefont
  {Yakhou}}, \bibinfo {author} {\bibfnamefont {V.}~\bibnamefont {Plakhty}},
  \bibinfo {author} {\bibfnamefont {H.}~\bibnamefont {Suzuki}}, \bibinfo
  {author} {\bibfnamefont {S.}~\bibnamefont {Gavrilov}}, \bibinfo {author}
  {\bibfnamefont {P.}~\bibnamefont {Burlet}}, \bibinfo {author} {\bibfnamefont
  {L.}~\bibnamefont {Paolasini}}, \bibinfo {author} {\bibfnamefont
  {C.}~\bibnamefont {Vettier}}, \ and\ \bibinfo {author} {\bibfnamefont
  {S.}~\bibnamefont {Kunii}},\ }\href {\doibase DOI:
  10.1016/S0375-9601(01)00338-3} {\bibfield  {journal} {\bibinfo  {journal}
  {Physics Letters A}\ }\textbf {\bibinfo {volume} {285}},\ \bibinfo {pages}
  {191 } (\bibinfo {year} {2001})}\BibitemShut {NoStop}%
\bibitem [{\citenamefont {Matsumura}\ \emph {et~al.}(2009)\citenamefont
  {Matsumura}, \citenamefont {Yonemura}, \citenamefont {Kunimori},
  \citenamefont {Sera},\ and\ \citenamefont {Iga}}]{Matsumura2009}%
  \BibitemOpen
  \bibfield  {author} {\bibinfo {author} {\bibfnamefont {T.}~\bibnamefont
  {Matsumura}}, \bibinfo {author} {\bibfnamefont {T.}~\bibnamefont {Yonemura}},
  \bibinfo {author} {\bibfnamefont {K.}~\bibnamefont {Kunimori}}, \bibinfo
  {author} {\bibfnamefont {M.}~\bibnamefont {Sera}}, \ and\ \bibinfo {author}
  {\bibfnamefont {F.}~\bibnamefont {Iga}},\ }\href {\doibase
  10.1103/PhysRevLett.103.017203} {\bibfield  {journal} {\bibinfo  {journal}
  {Phys. Rev. Lett.}\ }\textbf {\bibinfo {volume} {103}},\ \bibinfo {pages}
  {017203} (\bibinfo {year} {2009})}\BibitemShut {NoStop}%
\bibitem [{\citenamefont {Tanaka}\ \emph {et~al.}(2004)\citenamefont {Tanaka},
  \citenamefont {Staub}, \citenamefont {Katsumata}, \citenamefont {Lovesey},
  \citenamefont {Lorenzo}, \citenamefont {Narumi}, \citenamefont {Scagnoli},
  \citenamefont {Shimomura}, \citenamefont {Tabata}, \citenamefont {Onuki},
  \citenamefont {Kuramoto}, \citenamefont {Kikkawa}, \citenamefont {Ishikawa},\
  and\ \citenamefont {Kitamura}}]{Tanaka2004}%
  \BibitemOpen
  \bibfield  {author} {\bibinfo {author} {\bibfnamefont {Y.}~\bibnamefont
  {Tanaka}}, \bibinfo {author} {\bibfnamefont {U.}~\bibnamefont {Staub}},
  \bibinfo {author} {\bibfnamefont {K.}~\bibnamefont {Katsumata}}, \bibinfo
  {author} {\bibfnamefont {S.~W.}\ \bibnamefont {Lovesey}}, \bibinfo {author}
  {\bibfnamefont {J.~E.}\ \bibnamefont {Lorenzo}}, \bibinfo {author}
  {\bibfnamefont {Y.}~\bibnamefont {Narumi}}, \bibinfo {author} {\bibfnamefont
  {V.}~\bibnamefont {Scagnoli}}, \bibinfo {author} {\bibfnamefont
  {S.}~\bibnamefont {Shimomura}}, \bibinfo {author} {\bibfnamefont
  {Y.}~\bibnamefont {Tabata}}, \bibinfo {author} {\bibfnamefont
  {Y.}~\bibnamefont {Onuki}}, \bibinfo {author} {\bibfnamefont
  {Y.}~\bibnamefont {Kuramoto}}, \bibinfo {author} {\bibfnamefont
  {A.}~\bibnamefont {Kikkawa}}, \bibinfo {author} {\bibfnamefont
  {T.}~\bibnamefont {Ishikawa}}, \ and\ \bibinfo {author} {\bibfnamefont
  {H.}~\bibnamefont {Kitamura}},\ }\href
  {http://stacks.iop.org/0295-5075/68/i=5/a=671} {\bibfield  {journal}
  {\bibinfo  {journal} {EPL (Europhysics Letters)}\ }\textbf {\bibinfo {volume}
  {68}},\ \bibinfo {pages} {671} (\bibinfo {year} {2004})}\BibitemShut
  {NoStop}%
\bibitem [{\citenamefont {Zirngiebl}\ \emph {et~al.}(1984)\citenamefont
  {Zirngiebl}, \citenamefont {Hillebrands}, \citenamefont {Blumenr\"oder},
  \citenamefont {G\"untherodt}, \citenamefont {Loewenhaupt}, \citenamefont
  {Carpenter}, \citenamefont {Winzer},\ and\ \citenamefont
  {Fisk}}]{Zirngiebl1984}%
  \BibitemOpen
  \bibfield  {author} {\bibinfo {author} {\bibfnamefont {E.}~\bibnamefont
  {Zirngiebl}}, \bibinfo {author} {\bibfnamefont {B.}~\bibnamefont
  {Hillebrands}}, \bibinfo {author} {\bibfnamefont {S.}~\bibnamefont
  {Blumenr\"oder}}, \bibinfo {author} {\bibfnamefont {G.}~\bibnamefont
  {G\"untherodt}}, \bibinfo {author} {\bibfnamefont {M.}~\bibnamefont
  {Loewenhaupt}}, \bibinfo {author} {\bibfnamefont {J.~M.}\ \bibnamefont
  {Carpenter}}, \bibinfo {author} {\bibfnamefont {K.}~\bibnamefont {Winzer}}, \
  and\ \bibinfo {author} {\bibfnamefont {Z.}~\bibnamefont {Fisk}},\ }\href
  {\doibase 10.1103/PhysRevB.30.4052} {\bibfield  {journal} {\bibinfo
  {journal} {Phys. Rev. B}\ }\textbf {\bibinfo {volume} {30}},\ \bibinfo
  {pages} {4052} (\bibinfo {year} {1984})}\BibitemShut {NoStop}%
\bibitem [{\citenamefont {L{\"u}thi}\ \emph {et~al.}(1984)\citenamefont
  {L{\"u}thi}, \citenamefont {Blumenr{\"o}der}, \citenamefont {Hillebrands},
  \citenamefont {Zirngiebl}, \citenamefont {G{\"u}ntherodt},\ and\
  \citenamefont {Winzer}}]{Luthi1984}%
  \BibitemOpen
  \bibfield  {author} {\bibinfo {author} {\bibfnamefont {B.}~\bibnamefont
  {L{\"u}thi}}, \bibinfo {author} {\bibfnamefont {S.}~\bibnamefont
  {Blumenr{\"o}der}}, \bibinfo {author} {\bibfnamefont {B.}~\bibnamefont
  {Hillebrands}}, \bibinfo {author} {\bibfnamefont {E.}~\bibnamefont
  {Zirngiebl}}, \bibinfo {author} {\bibfnamefont {G.}~\bibnamefont
  {G{\"u}ntherodt}}, \ and\ \bibinfo {author} {\bibfnamefont {K.}~\bibnamefont
  {Winzer}},\ }\href {http://dx.doi.org/10.1007/BF01469434} {\bibfield
  {journal} {\bibinfo  {journal} {Zeitschrift f{\"u}r Physik B Condensed
  Matter}\ }\textbf {\bibinfo {volume} {58}},\ \bibinfo {pages} {31} (\bibinfo
  {year} {1984})},\ \bibinfo {note} {10.1007/BF01469434}\BibitemShut {NoStop}%
\bibitem [{\citenamefont {Sakai}\ \emph {et~al.}(1997)\citenamefont {Sakai},
  \citenamefont {Shiina}, \citenamefont {Shiba},\ and\ \citenamefont
  {Thalmeier}}]{Sakai1997}%
  \BibitemOpen
  \bibfield  {author} {\bibinfo {author} {\bibfnamefont {O.}~\bibnamefont
  {Sakai}}, \bibinfo {author} {\bibfnamefont {R.}~\bibnamefont {Shiina}},
  \bibinfo {author} {\bibfnamefont {H.}~\bibnamefont {Shiba}}, \ and\ \bibinfo
  {author} {\bibfnamefont {P.}~\bibnamefont {Thalmeier}},\ }\href {\doibase
  10.1143/JPSJ.66.3005} {\bibfield  {journal} {\bibinfo  {journal} {Journal of
  the Physical Society of Japan}\ }\textbf {\bibinfo {volume} {66}},\ \bibinfo
  {pages} {3005} (\bibinfo {year} {1997})}\BibitemShut {NoStop}%
\bibitem [{\citenamefont {Plakhty}\ \emph {et~al.}(2005)\citenamefont
  {Plakhty}, \citenamefont {Regnault}, \citenamefont {Goltsev}, \citenamefont
  {Gavrilov}, \citenamefont {Yakhou}, \citenamefont {Flouquet}, \citenamefont
  {Vettier},\ and\ \citenamefont {Kunii}}]{Plakhty2005}%
  \BibitemOpen
  \bibfield  {author} {\bibinfo {author} {\bibfnamefont {V.~P.}\ \bibnamefont
  {Plakhty}}, \bibinfo {author} {\bibfnamefont {L.~P.}\ \bibnamefont
  {Regnault}}, \bibinfo {author} {\bibfnamefont {A.~V.}\ \bibnamefont
  {Goltsev}}, \bibinfo {author} {\bibfnamefont {S.~V.}\ \bibnamefont
  {Gavrilov}}, \bibinfo {author} {\bibfnamefont {F.}~\bibnamefont {Yakhou}},
  \bibinfo {author} {\bibfnamefont {J.}~\bibnamefont {Flouquet}}, \bibinfo
  {author} {\bibfnamefont {C.}~\bibnamefont {Vettier}}, \ and\ \bibinfo
  {author} {\bibfnamefont {S.}~\bibnamefont {Kunii}},\ }\href {\doibase
  10.1103/PhysRevB.71.100407} {\bibfield  {journal} {\bibinfo  {journal} {Phys.
  Rev. B}\ }\textbf {\bibinfo {volume} {71}},\ \bibinfo {pages} {100407}
  (\bibinfo {year} {2005})}\BibitemShut {NoStop}%
\bibitem [{\citenamefont {Arrott}(1957)}]{Arrott1957}%
  \BibitemOpen
  \bibfield  {author} {\bibinfo {author} {\bibfnamefont {A.}~\bibnamefont
  {Arrott}},\ }\href {\doibase 10.1103/PhysRev.108.1394} {\bibfield  {journal}
  {\bibinfo  {journal} {Phys. Rev.}\ }\textbf {\bibinfo {volume} {108}},\
  \bibinfo {pages} {1394} (\bibinfo {year} {1957})}\BibitemShut {NoStop}%
\bibitem [{\citenamefont {Amara}\ \emph {et~al.}(2010)\citenamefont {Amara},
  \citenamefont {Gal\'era}, \citenamefont {Aviani},\ and\ \citenamefont
  {Givord}}]{Amara2010}%
  \BibitemOpen
  \bibfield  {author} {\bibinfo {author} {\bibfnamefont {M.}~\bibnamefont
  {Amara}}, \bibinfo {author} {\bibfnamefont {R.-M.}\ \bibnamefont {Gal\'era}},
  \bibinfo {author} {\bibfnamefont {I.}~\bibnamefont {Aviani}}, \ and\ \bibinfo
  {author} {\bibfnamefont {F.}~\bibnamefont {Givord}},\ }\href {\doibase
  10.1103/PhysRevB.82.224411} {\bibfield  {journal} {\bibinfo  {journal} {Phys.
  Rev. B}\ }\textbf {\bibinfo {volume} {82}},\ \bibinfo {pages} {224411}
  (\bibinfo {year} {2010})}\BibitemShut {NoStop}%
\bibitem [{\citenamefont {Callen}\ and\ \citenamefont
  {Callen}(1963)}]{Callen1963}%
  \BibitemOpen
  \bibfield  {author} {\bibinfo {author} {\bibfnamefont {E.~R.}\ \bibnamefont
  {Callen}}\ and\ \bibinfo {author} {\bibfnamefont {H.~B.}\ \bibnamefont
  {Callen}},\ }\href {\doibase 10.1103/PhysRev.129.578} {\bibfield  {journal}
  {\bibinfo  {journal} {Phys. Rev.}\ }\textbf {\bibinfo {volume} {129}},\
  \bibinfo {pages} {578} (\bibinfo {year} {1963})}\BibitemShut {NoStop}%
\bibitem [{\citenamefont {Sera}\ \emph {et~al.}(1988)\citenamefont {Sera},
  \citenamefont {Sato},\ and\ \citenamefont {Kasuya}}]{Sera1988}%
  \BibitemOpen
  \bibfield  {author} {\bibinfo {author} {\bibfnamefont {M.}~\bibnamefont
  {Sera}}, \bibinfo {author} {\bibfnamefont {N.}~\bibnamefont {Sato}}, \ and\
  \bibinfo {author} {\bibfnamefont {T.}~\bibnamefont {Kasuya}},\ }\href
  {\doibase 10.1143/JPSJ.57.1412} {\bibfield  {journal} {\bibinfo  {journal}
  {Journal of the Physical Society of Japan}\ }\textbf {\bibinfo {volume}
  {57}},\ \bibinfo {pages} {1412} (\bibinfo {year} {1988})}\BibitemShut
  {NoStop}%
\bibitem [{\citenamefont {Morin}\ \emph {et~al.}(1980)\citenamefont {Morin},
  \citenamefont {Schmitt},\ and\ \citenamefont {de~Lacheisserie}}]{Morin1980A}%
  \BibitemOpen
  \bibfield  {author} {\bibinfo {author} {\bibfnamefont {P.}~\bibnamefont
  {Morin}}, \bibinfo {author} {\bibfnamefont {D.}~\bibnamefont {Schmitt}}, \
  and\ \bibinfo {author} {\bibfnamefont {E.}~\bibnamefont {de~Lacheisserie}},\
  }\href@noop {} {\bibfield  {journal} {\bibinfo  {journal} {Phys. Rev.}\
  }\textbf {\bibinfo {volume} {21}},\ \bibinfo {pages} {1742} (\bibinfo {year}
  {1980})}\BibitemShut {NoStop}%
\bibitem [{\citenamefont {Kasuya}(1998)}]{Kasu1998}%
  \BibitemOpen
  \bibfield  {author} {\bibinfo {author} {\bibfnamefont {T.}~\bibnamefont
  {Kasuya}},\ }\href@noop {} {\bibfield  {journal} {\bibinfo  {journal}
  {Journal of the Physical Society of Japan}\ }\textbf {\bibinfo {volume}
  {67}},\ \bibinfo {pages} {33} (\bibinfo {year} {1998})}\BibitemShut {NoStop}%
\end{thebibliography}
\end{document}